\documentclass[12pt]{iopart}
\usepackage{iopams}
\usepackage{graphicx}
\input amssym.def 
\input amssym

\begin{document}

\title[]{Distinguished three-qubit \lq magicity' via 
automorphisms of the split Cayley hexagon}

\author{ Michel Planat$^{1}$, Metod Saniga$^{2}$ and Fr\'ed\'eric Holweck$^{3}$}

\vspace*{.1cm}
\address{$^1$Institut FEMTO-ST, CNRS, 32 Avenue de
l'Observatoire, F-25044 Besan\c con, France }
\ead{michel.planat@femto-st.fr}

\vspace*{.2cm}

\address{$^2$Astronomical Institute, Slovak Academy of Sciences, SK-05960 Tatransk\'{a} Lomnica, Slovak Republic}
\ead{msaniga@astro.sk}

\vspace*{.2cm}

\address{$^3$Laboratoire M3M, Universit\'e de Technologie de Belfort-Montb\'eliard, F-90010 Belfort,
France}
\ead{frederic.holweck@utbm.fr}

\vspace*{.1cm}

\begin{abstract}
Disregarding the identity, the remaining 63 elements of the generalized three-qubit Pauli group are found to contain 12\,096 distinct copies of Mermin's magic pentagram. Remarkably, 12\,096 is also the number of automorphisms of the smallest split Cayley hexagon. We give a few solid arguments showing that this may not be a mere coincidence. These arguments are mainly tied to the structure of certain types of geometric hyperplanes of the hexagon. It is further demonstrated that also an $(18_{2}, 12_{3})$-type of magic configurations, recently proposed by Waegell and Aravind ({\it J. Phys. A: Math. Theor.} {\bf 45} (2012) 405301), seems to be intricately linked with automorphisms of the hexagon. Finally, the entanglement properties exhibited by edges of both pentagrams and these particular Waegell-Aravind configurations are addressed.

\end{abstract}

\pacs{03.65.Aa, 03.65.Fd, 03.65.Ud, 02.10.Ox}

\section{Introduction}
\noindent
Quantum contextuality can succinctly  be formulated as the non-existence of elements of physical reality in the absence of measurements and, when a selected set-up is used for a measurement, the spontaneous existence of all compatible ones.
There are several approaches to this paradoxical subject, these being mainly related to complementarity and non-locality \cite{Liang2011} and, to some extent, also to entanglement \cite{Greenberger90}. The simplest technical formulation seems to be the Bell-Kochen-Specker (BKS) theorem, the impossibility to assign all rays of a $n$-dimensional ($n \ge 3$) Hilbert space  binary values (0 for false, 1 for true) in such a way that exactly one ray in each complete basis is labeled by 1 \cite{Kochen1967,Bengtsson2012}.

Recently, a pentagram (or, a five-pointed star) has been used as an efficient archetype for dealing with some contexts. In particular, we have in mind three-dimensional Klyachko's proof of non-contextuality via the failure of transitivity of implications for counter-factual statements \cite{Liang2011}. In a slightly different direction, Mermin's parity proof of the BKS theorem relies on an appropriate labeling of the vertices of the pentagram, frequently bearing an adjective `magic', by three-qubit operators/observables \cite{Mermin1993,Planat2012}. As observed by Arkhipov \cite{Arkhipov2012}, the knowledge of all possible quantum realizations of the pentagram is worth studying and is precisely the main objective of our paper. Our approach may be compared to that of \cite{Ruuge2005}. In the latter work it was found that the $E_8$-root system is the sought-for saturated three-qubit ray configuration. Here, we shall further advance the ideas of \cite{Sanigaetal2012} and show that the finite geometry that seems to entail all essential features of three-qubit contextuality and associated magic configurations is indeed the split Cayley hexagon of order two (occasionally referred to as the $G_2(2)$-hexagon since $G_2(2)$ is its automorphism group) \cite{psm,Schroth1999}. 

The paper is organized as follows. In Sec.\,2, we describe two basic aggregates of three-qubit observables that allow an operator proof of the BKS theorem, namely the pentagram first proposed by Mermin \cite{Mermin1993} and an $(18_{2}, 12_{3})$-configuration recently found by Waegell and Aravind \cite{Aravind2012}, and introduce the concept of their type. In Sec.\,3, we first introduce  the split Cayley hexagon of order two and list basic properties of its geometric hyperplanes. Then, we deal with a type split of the whole set of  Mermin's pentagrams and that of WA's within and across various types of geometric hyperplanes of the hexagon. 
In Sec. 4, a brief appraisal of the GHZ type of entanglement exhibited by the edges of both magic configurations is given, again in relation to various types of hexagon's geometric hyperplanes. Finally, Sec. 5 lists a few notable configurations stemming from sets of pentagrams. 

The paper can also be regarded as a first attempt to correlate quantum contextuality and the GHZ-type of entanglement in the language of finite geometry.

\section{Magic pentagram and a magic Waegell-Aravind configuration}
\label{Sect2}
\noindent
The notation adopted is similar to that of \cite{Sanigaetal2012}. $X$, $Y$ and $Z$ are the Pauli spin matrices in direction $x$, $y$ and $z$, respectively, and $I$ is the identity matrix whose dimension depends on the context. Our three-qubit observables are tensor products from the set of the above-given matrices; we use for them a short-hand notation, e.\,g., $XIY \equiv X \otimes I \otimes Y$.

\subsection*{Magic pentagrams}

\begin{figure}
\centering
\includegraphics[width=7cm]{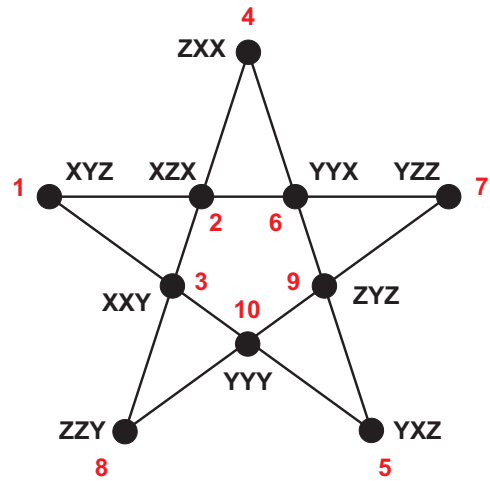}
\caption{A magic pentagram of type 1.}
\label{fig1}
\end{figure}

\begin{figure}
\centering
\includegraphics[width=15cm]{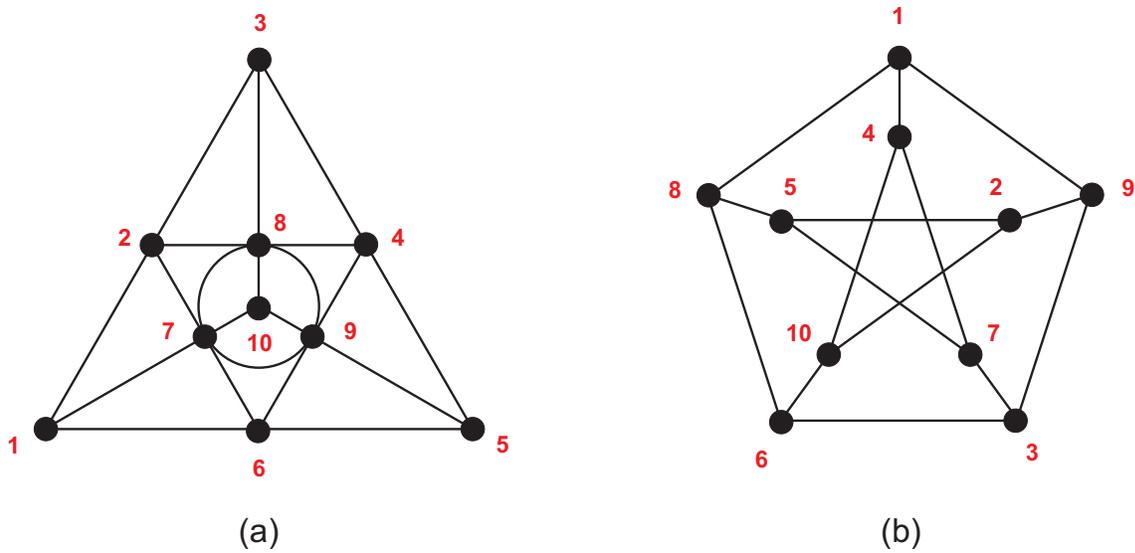}
\caption{(a) The non-realizable $10_3$-configuration and (b) the complement of its collinearity graph, the Petersen graph.}
\label{fig2}
\end{figure}

A magic pentagram comprises five sets (edges) of four mutually commuting three-qubit observables (vertices). These sets share pairwise a single element and the product of observables in each of them yields either $+I$ or  $-I$, with the understanding that the latter case occurs an odd number of times. Thus, there are tree types of three-qubit magic pentagrams according as all the products (type 1), three products (type 2) or just one product (type 3) are/is $-I$.
Fig. \ref{fig1} depicts an example of a pentagram of type 1. It is worth mentioning here that the pentagram can be also redrawn as the unique `non-realizable' $10_3$-configuration\footnote{It is obvious that the product of the three operators located on a line of the $10_3$-configuration is not equal to $\pm I$.} \cite{Gropp1997}, as shown in Fig. \ref{fig2}{\it a}.  This particular $10_3$-configuration is  remarkable in that the complement of its collinearity graph is nothing but the famous Petersen graph, shown in Fig. \ref{fig2}{\it b}. 

As a slight digression from the main theme, we mention that the pentagram graph and, similarly, the collinearity graph of the non-realizable $10_3$-configuration, are `non-trivial'  in terms of their Shannon capacity  \cite{Bekius2011}. Let us denote $G\square H$, $G \times H$ and $G \boxtimes H=(G\square H)\cup(G \times H)$ the cartesian product, the tensor product and the strong product of graphs $G$ and $H$, respectively. The Shannon capacity $\Theta(G)$ of the graph $G$ is the maximum number of $k$-letter messages that can be sent through a channel without a risk of confusion. It is defined by the expression
\begin{equation*}
\Theta(G)=\mbox{sup}_k\sqrt[k]{\alpha(G^k)}.
\label{Shan}
\end{equation*}
where $G^k=G \boxtimes G \boxtimes \cdots \boxtimes G $ (with $k$ terms in the product) and  $\alpha(G)$ is the size of a maximum independent set (also called the independence number) of $G$. For a general graph one has the bound $\Theta(G)\ge \alpha(G)$. For a perfect graph the bound is reached, otherwise the graph is said to be non-trivial.
For the pentagram graph one numerically gets $\Theta(G)\ge \sqrt{5}>\alpha(G)=2$ (just taking the product $G^2$ for approximating the Shannon capacity of $G$). Note that the pentagon graph has the tight bound $\sqrt{5}$ for its Shannon capacity.

\subsection*{Waegell-Aravind magic configurations}

\begin{figure}[h]
\centering
\includegraphics[width=7cm]{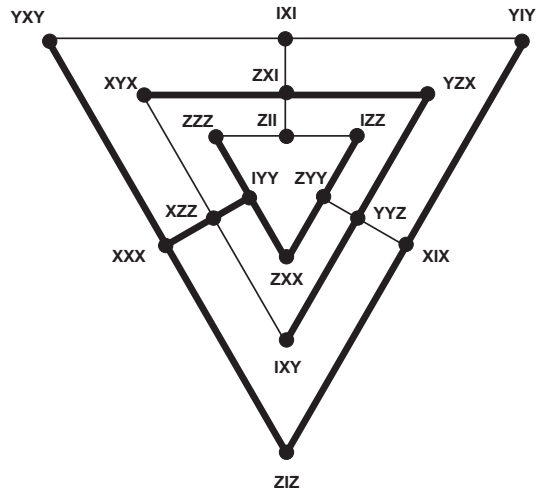}
\caption{A WA-configuration of type 1; thin (resp. thick) lines have the product of their operators equal to $+I$ (resp. $-I$).}
\label{fig3}
\end{figure}

Recently, a wealth of other magic configurations have been discovered \cite{Aravind2012}. One of them, denoted as $(18_2,12_3)$ and referred to as a WA-configuration in the sequel,  contains $18$ vertices (observables) and 12 edges, each vertex (observable) being on 2 edges and each edge comprising $3$ vertices (observables). The product of observables along each edge is $+I$ or $-I$, with the latter possibility occurring an odd number of times. Performing an exhaustive computer search we have found that they are of four distinct types (1 to 4) according as the number of edges yielding $-I$ is 7, 5, 3 or 1, respectively. The WA-configuration given in \cite[Fig. 5]{Aravind2012} is of type 3; Fig. \ref{fig3} illustrates a WA-configuration of type 1.

\section{Three-qubit \lq magicity' and the smallest split Cayley hexagon}
\label{Sect3}
\noindent

\subsection*{The hexagon, its symplectic embeddings and geometric hyperplanes}
\begin{figure}[h]
\centering
\includegraphics[width=9.2cm]{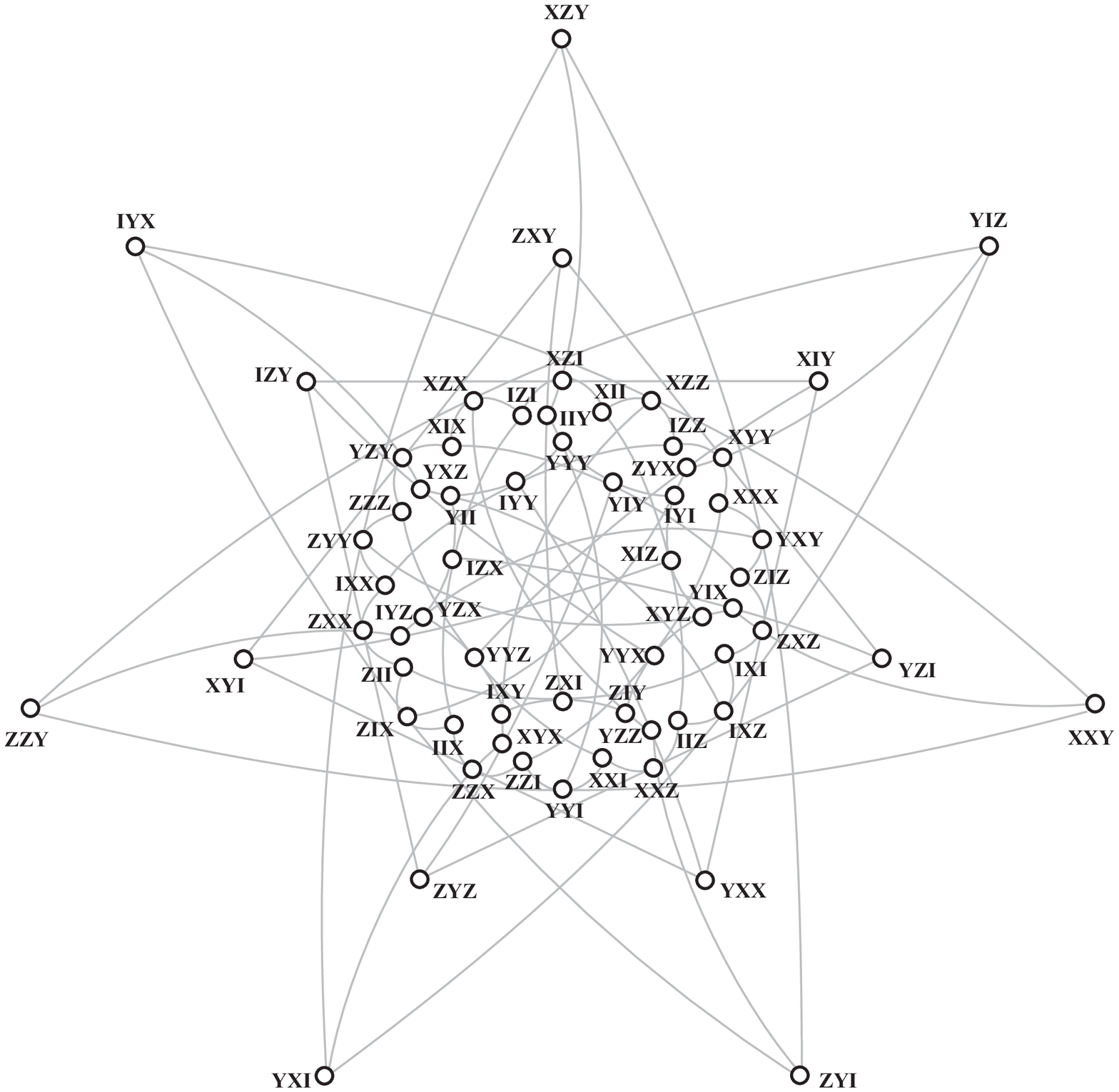}\includegraphics[width=9.2cm]{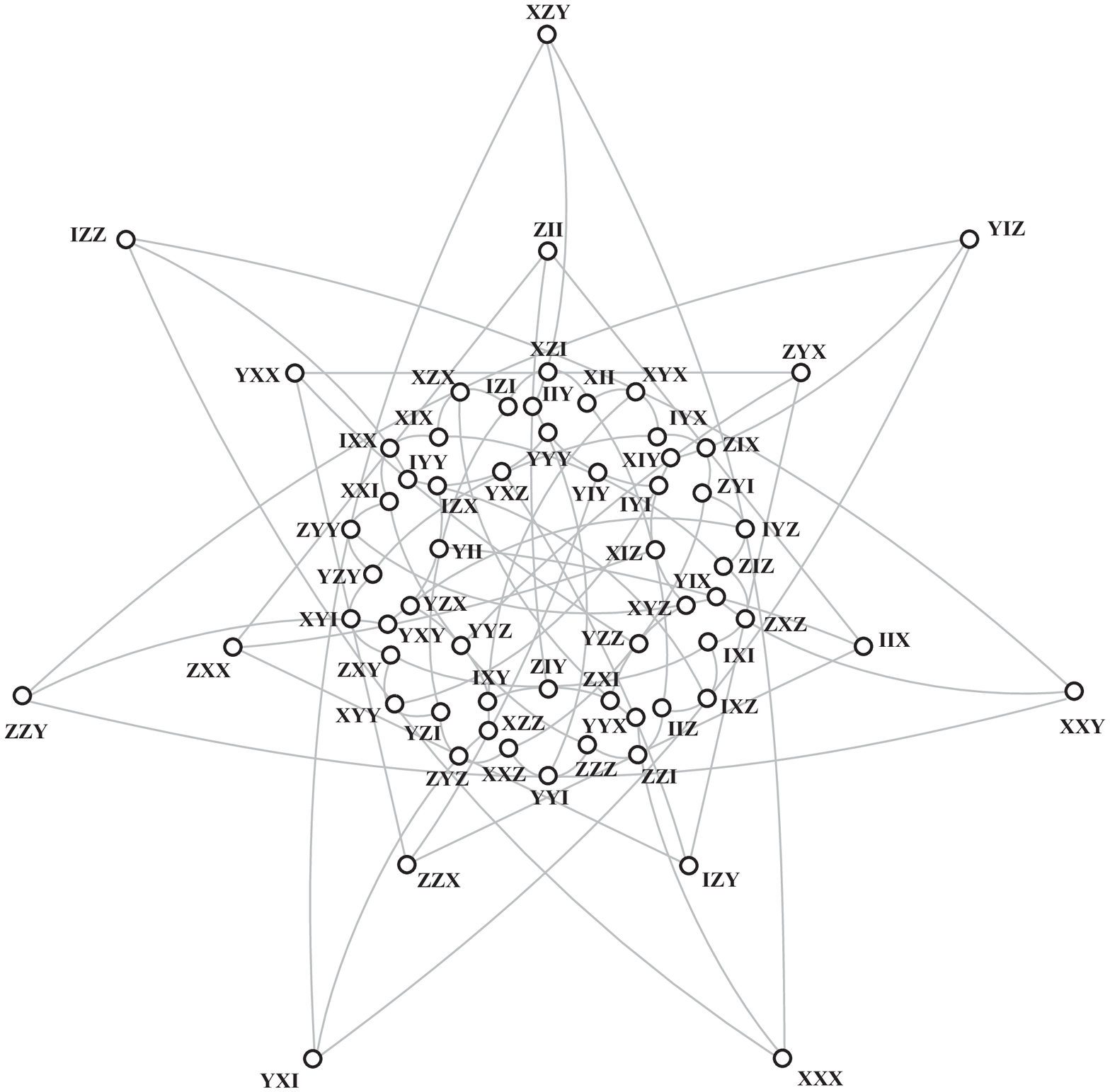}
\caption{A classical ({\it left}) and skew ({\it right}) symplectic embedding of the hexagon in terms of the elements of the three-qubit Pauli group. The points of the hexagon are represented by small circles and its lines by straight segments and/or arcs joining three circles each (based on the drawings given in \cite{psm,Schroth1999}).}
\label{fig4}
\end{figure}

A split Cayley hexagon of order two, in what follows simply the hexagon, is a point-line incidence geometry whose incidence graph has girth 12 and diameter 6 such that there are exactly three points incident with any line and three lines incident with any point, and which contains a copy of the Heawood graph --- the incidence graph of the Fano plane; from the definition it follows that  the hexagon possesses 63 points and the same number of lines (see, for example, \cite{psm,Schroth1999}). A remarkable, and essential for our further considerations, property of our hexagon is that it can be embedded into the symplectic polar space $W(5, 2)$ --- the space underlying the commutation properties of the 63 non-trivial elements of the generalized three-qubit Pauli group (see, for example, \cite{Sanigaetal2012,SanigaPlanat2007, Planat2011} and references therein). Moreover, it has recently been shown \cite{Coolsaet} that there are two such symplectic embeddings; a more symmetric one, called classical and a less symmetric one, termed skew.  An example of either of them is portrayed in Fig. \ref{fig4}. In what follows we shall exclusively be dealing with the former type of embedding; and this not only because of its symmetry-appeal, but also due to the fact that it has already been employed in a couple of papers concerned with important physical applications of the hexagon \cite{Sanigaetal2012,Levay2008}.

\begin{table}[t]
\begin{center}
\small
\begin{tabular}{||l|l|cccr|l|l||}
\hline \hline
Class & FJ Type & Pts & Lns & DPts & Cps & StGr & Cmp \\
\hline\hline
\vspace*{-.30cm}
&&&&&&&\\
I & $\cal V$$_{2}$(21;21,0,0,0) & 21 & 0 & 0 & 36 & $PGL(2,7)$ & Heawood $\cup$ Coxeter\\
\hline
II& $\cal V$$_{7}$(23;16,6,0,1) &23& 3 & 1 & 126 & $(4 \times 4):S_3$  & \\
\hline
III& $\cal V$$_{11}$(25;10,12,3,0) &25&6&0&504& $S_4$ & \\
\hline
IV & $\cal V$$_{1}$(27;0,27,0,0) & 27 & 9 & 0 & 28 & $X_{27}^{+}:QD_{16}$ & Pappus $\cup$ Pappus\\
& $\cal V$$_{8}$(27;8,15,0,4) & 27 &  9 & 3+1 & 252 & $2 \times S_4$ & \\
& $\cal V$$_{13}$(27;8,11,8,0)  & 27 & 9 & 0 & 756 & $D_{16}$ & \\
& $\cal V$$_{17}$(27;6,15,6,0) & 27 & 9 &  0  & 1008 & $D_{12}$ & \\
\hline
V & $\cal V$$_{12}$(29;7,12,6,4) & 29 & 12 &  4  & 504 & $S_4$ &  \\
& $\cal V$$_{18}$(29;5,12,12,0) & 29 & 12 & 0 & 1008 & $D_{12}$ & \\
& $\cal V$$_{19}$(29;6,12,9,2) & 29 & 12 &  2nc  & 1008 & $D_{12}$  & \\
&  $\cal V$$_{23}$(29;4,16,7,2) & 29 & 12 &  2c  & 1512 & $D_{8}$ & \\
\hline
VI & $\cal V$$_{6}$(31;0,24,0,7) & 31 & 15 & 6+1 & 63 & $(4 \times 4):D_{12}$ & Dyck \\
& $\cal V$$_{24}$(31;4,12,12,3) & 31 & 15 & 2+1 & 1512 & $D_8$ & \\
& $\cal V$$_{25}$(31;4,12,12,3) & 31 & 15 & 3 & 2016 & $S_3$ & \\
\hline
VII & $\cal V$$_{14}$(33;4,8,17,4) & 33 & 18 & 2+2 & 756 & $D_{16}$ & \\
& $\cal V$$_{20}$(33;2,12,15,4) & 33 & 18 & 3+1 & 1008 & $D_{12}$ & \\
\hline
VIII& $\cal V$$_{3}$(35;0,21,0,14) & 35 & 21 & 14 & 36 & $PGL(2,7)$ & Coxeter \\
& $\cal V$$_{16}$(35;0,13,16,6) & 35 & 21 & 4+2 & 756 & $D_{16}$ & \\
& $\cal V$$_{21}$(35;2,9,18,6) & 35 & 21 &  6  & 1008 & $D_{12}$ & \\
\hline
IX & $\cal V$$_{15}$(37;1,8,20,8) & 37 & 24 & 8 & 756 & $D_{16}$ & \\
& $\cal V$$_{22}$(37;0,12,15,10) & 37 & 24 &6+3+1& 1008 & $D_{12}$ & \\
\hline
X & $\cal V$$_{10}$(39;0,10,16,13) & 39 & 27 & 8+4+1 & 378 & $8:2:2$ & \\
\hline
XI& $\cal V$$_{9}$(43;0,3,24,16)  &43&33&12+3+1&252&$2 \times S_4$ & non-realizable $10_3$\\
\hline
XII& $\cal V$$_{5}$(45;0,0,27,18)  &45&36&18&56&$X_{27}^{+}:D_{8}$  & Pappus \\
\hline
XIII& $\cal V$$_{4}$(49;0,0,21,28) & 49 & 42 & 28 & 36 & $PGL(2,7)$ & Heawood \\
\hline \hline
\end{tabular}
\vspace*{0.4cm}
\caption{A classification of the geometric hyperplanes of the 
hexagon. Each type is characterized by the size of its point- (`Pts') and line- (`Lns') sets, number of deep points (`DPts'), total number of distinct copies (`Cps') and the stabilizer group (`StGr') of its orbit; for some types we also list the cubic graph isomorphic to the complement (`Cmp') of a copy of the hyperplane.}
\end{center}
\end{table}

Another distinguished feature of the hexagon is that it contains a relatively large number of distinct types of geometric hyperplanes. A geometric hyperplane of a point-line incidence structure is a subset of the point-set such that every line of the structure either belongs  fully to the subset, or shares with it just a single point \cite{Ronan};
a point of a geometric hyperplane is called deep if all the lines passing through it are fully contained in the hyperplane. A total of $2^{14} - 1 = 16\,383$ geometric hyperplanes of the hexagon were fully classified in \cite{Frohard1994}. They fall into 25 distinct types (according to the orbits of its automorphism group) and 13 classes (in terms of the sizes of their point-/line-sets). This classification --- adopted, with a slight modification, from \cite{Sanigaetal2012} --- is given in Table 1. This table also features the compact Frohardt-Johnson `five-tuple' notation \cite{Frohard1994}, ${\cal V}_{k}(n; n_0, n_1, n_2, n_3)$, meaning that a hyperplane of the $k$-th type, $1 \leq k \leq 25$, has $n$ points of which $n_s$, $s \in \{0, 1, 2, 3\}$, belong to exactly $s$ lines contained in the hyperplane; obviously, $n_3$ is the number of deep points of a hyperplane. 

Geometric hyperplanes, apart from being a key ingredient of our subsequent reasoning, also lend themselves as a useful tool to tell apart the two symplectic embeddings of the hexagon. For given any of the 63 elements of the three-qubit Pauli group, there are 31 elements (the element itself inclusive) that commute with it and they always form a geometric hyperplane of the hexagon. The difference between the two embeddings lies with the fact that whereas for a {\it classical} embedding this hyperplane is for each element of {\it the same} type, namely ${\cal V}_6$, for a {\it skew} embedding it is of {\it two} different kinds; for 15 elements being of type ${\cal V}_6$ and for the remaining 48 elements of type ${\cal V}_{24}$.

\subsection*{Magic pentagrams and geometric hyperplanes of the hexagon}

Our next task is to find all magic pentagrams living in $W(5,2)$. To this end, we employ the property that an edge of any such pentagram represents an affine plane of order two, AG$(2,3)$ \cite{SanLev2012}.
Since $W(5,2)$ contains 135 Fano planes, and each Fano plane gives rise to seven copies of AG$(2,3)$, this symplectic space features altogether 945 copies of AG$(2,3)$.
Next, we search for quintuples of these such that any two of them share a single point. At the last stage, we keep only those quintuples, where the product of observables yielding $-I$ occurs an odd number of times. With the help of computer, we arrive at a total of 12\,096 magic pentagrams, with the following distribution of their types (one to three, respectively):
\begin{equation}
12096 = 108 + 4104  + 7884   = 108 (1 + 38 + 73).
\end{equation}
To get further insight into the structure of the set of pentagrams, we looked for those that are, as point-sets, located in a randomly-chosen copy of geometric hyperplane of a given type. The results of our computer analysis are summarized in Table 2, which reveals a number of very interesting facts. First, we see that the smallest hyperplane to feature a pentagram has 29 points and is of type $\mathcal{V}_{12}$. Next, comparing Tables 1 and 2, one observes that only classes VII to XIII are fully represented.  However, by far the most intriguing are observations that one gets when making the product of the number of pentagrams (`Pents') in a given hyperplane's copy with the number of copies  (`Cps') of the hyperplane in the hexagon, as given in the last column of Table 2.  One sees a prevailing number of integer multiples of the number $12 096$ --- the number identical to the order of $G_2(2)$, the group of automorphisms of our hexagon! Then, we have two multiples of the number  $20 160$, this being the order of $A_8$, and, finally, there is a single multiple of the number $336$, the order of group $SL(2, 7)$. Also this latter case is quite interesting, for it is associated with the hyperplane of type $\mathcal{V}_{4}$; a hyperplane of this type, apart from being of the largest possible size, is also remarkable by the fact that its complement is nothing but the incidence graph of the Fano plane (the Heawood graph --- see Table 1).

\begin{table}[ht]
\begin{center}
\small
\begin{tabular}{||l|r|r|r|r||}
\hline \hline
 Hyperplane & Pts & Cps & Pents &  Cps $\times$ Pents \\
\hline \hline
$\mathcal{V}_{12}$ & 29 &  $504$               &  $24=1+5+18$   &  $12096$ \\
\hline
$\mathcal{V}_{24}$ & 31 & $1512$       &  $8=2(1+3+0)$    &  $12096$ \\
\hline
$\mathcal{V}_{20}$ & 33 & $1008$       &  $20=0+0+20$    &  $20160$ \\
\hline
$\mathcal{V}_{14}$ & 33 & $756$       &  $16=2(0+1+7)$    &  $12096$ \\
\hline
$\mathcal{V}_{3}$  & 35 & $36$       &  $336=2(1+42+125)$    &  $12096$ \\
\hline
$\mathcal{V}_{16}$ & 35 & $756$       &  $32=2(0+3+13)$    &  $2 \times12096$ \\
\hline
$\mathcal{V}_{21}$ & 35 & $1008$       &  $36=12(0+1+2)$    &  $3 \times12096$ \\
\hline
$\mathcal{V}_{15}$ & 37 & $756$       &  $96=2(1+11+36)$    &  $6 \times12096$ \\
\hline
$\mathcal{V}_{22}$ & 37 & $1008$       &  $48=0+0+48$    &  $4 \times12096$ \\
$\mathcal{V}_{22}$ (2-nd copy)  &  &     &  $48=0+26+22$    &   \\
\hline
$\mathcal{V}_{10}$ & 39 & $378$       &  $160=4(0+9+31)$    &  $3 \times 20160$ \\
\hline
$\mathcal{V}_{9}$  & 43 &  $252$       &  $336=8(1+19+22)$    &  $7 \times 12096$ \\
\hline
$\mathcal{V}_{5}$  & 45 &  $56$       &  $432=8(2+17+35)$    &  $2 \times 12096$ \\
\hline
$\mathcal{V}_{4}$ & 49 & $36$       &  $1456=2(16+273+439)$    &  $156 \times 336$~~~ \\
\hline
trivial   & 63 & $1$   & $12096=108(1+38+73)$ &   $12096$ \\
\hline \hline
\end{tabular}
\label{table2}
\vspace*{0.4cm}
\caption{Occurrence of magic pentagrams of different types inside a selected copy of the hyperplane of a given type. An example of $\mathcal{V}_{22}$ shows that although the total number of pentagrams remains the same,  the ratio between their types may change when one takes a different copy of the hyperplane of the same type. A copy selected for $\mathcal{V}_{3}$ consists of all 35 symmetric elements. No magic pentagrams were found in the remaining types of hyperplanes.}
\end{center}
\end{table}

\subsection*{WA-configurations and geometric hyperplanes of the hexagon}
Our computer search for WA-configurations followed the strategy similar to that of the preceding subsection. A closer look at Fig.\,3  reveals that a WA-configuration consists of 18 observables equally distributed into three disjoint, concentric triangles, the triangles being glued together through `midpoints' of their sides by three edges. In the language of $W(5,2)$, these triangles correspond to nothing but  three disjoint copies of a punctured Fano plane, with one additional line deleted --- namely that joining the three midpoints in each triangle. After finding such an aggregate of observables, it only remained to check for and retain those where the number of `negative edges' was odd. Although, unfortunately, our computer power was not enough to find all WA-configurations in $W(5, 2)$, we luckily could perform this task when restricting to geometric hyperplanes and the corresponding results are listed in Table 3. There are, as expected, many less types of hyperplanes when compared with the previous case. Yet, we still encounter the three distinguished numbers of the preceding section, although two of them go here also as fractional multiples. It is also worth noting that  the hyperplane of type $\mathcal{V}_{3}$ is absent in Table 3; hence, there are no WA-configurations in the set of symmetric elements. The number $40320$ on the last line of the table is our conjectured estimate of the total number of WA-configurations in $W(5, 2)$, based on a chain of reasoning given in the following subsection.

\begin{table}[ht]
\begin{center}
\small
\begin{tabular}{||l|r|r|r|r||}
\hline \hline
Hyperplane & Pts & Cps &  WAs &  Cps $\times$ WAs \\
\hline \hline
$\mathcal{V}_{22}$ & 37 & $1008$       &  $7=0+0+1+6$    &  $21 \times 336$~~~ \\
$\mathcal{V}_{22}$ (2-nd copy) &  &       &  $7=0+5+1+1$    &   \\
\hline
$\mathcal{V}_{10}$ & 39 & $378$       &  $16=4(0+0+1+3)$    &  $1/2 \times 12096$ \\
\hline
$\mathcal{V}_{9}$  & 43 &  $252$       &  $40=2(0+11+8+1)$    &  $1/2 \times 20160$ \\
\hline
$\mathcal{V}_{5}$  & 45 &  $56$       &  $126=2(7+32+21+3)$    &  $21 \times 336$~~~ \\
\hline
$\mathcal{V}_{4}$  & 49 & $36$       &  $336=12(5+17+5+1)$    &  $12096$ \\
\hline
trivial    & 63 & $1$   & $40320$(?) &  $40320$(?)  \\
\hline \hline
\end{tabular}
\label{table3}
\vspace*{0.4cm}
\caption{Occurrence of magic WA-configurations of different types inside a selected copy of the hyperplane of a given type. An example of $\mathcal{V}_{22}$ shows that although the total number of WAs remains the same,  the ratio between their types may change when one takes a different copy of the hyperplane of the same type. No WA-configurations were found in the remaining types of hyperplanes.}
\end{center}
\end{table}

\subsection*{Group-theoretical justification of the distinguished numbers}
Let us try to understand the occurrence of the three remarkable numbers in the language of group theory, starting with the smallest of them. To this end, let us take the set of 35 symmetric elements/observables.
As already mentioned (see also \cite{psm}), these form in the hexagon a geometric hyperplane of type $\mathcal{V}_{3}$, which has 21 lines (see Table 1). In the full space $W(5,2)$ these 35 elements occupy as many as 105 lines. Now, consider the point-line incidence structure formed by these 35 points and 105 lines; its group of automorphims is isomorphic to $S_8$, of order $40320$. Next, a magic pentagram features 120 automorphisms, forming the group  isomorphic to $S_5$. We see that $|S_8|/|S_5|$ = 40320/120 = 336, the number of magic pentagrams in the set of symmetric elements! Similarly, taking the whole set of 63 elements and 315 lines these form in $W(5,2)$, the resulting structure features 1451520 automorphisms, comprising the group isomorphic to  $Sp(6,2)$; in this case we find, as expected, $|Sp(6,2)|/|S_5|$ = 1451520/120 = 12096, the total number of pentagrams. In the same vein, given the fact that the group of automorphism of a magic WA-configuration is isomorphic to $S_3^2$, we have  $|Sp(6,2)|/|S_3^2|$ = 1451520/36 = 40320, which is conjectured to be the total number of such configurations in $W(5,2)$.

\section{Entanglement in pentagrams and WA-configurations}
\label{Sect4}
As each edge, of both a pentagram and a WA-configuration, features a set of {\it mutually} commuting operators, it can also be associated with a set of eigenvectors common to the operators in question. Hence, a natural question arises: what kind of entanglement are such vectors endowed  with?
It is well known (see, e.\,g., \cite{Coffman2000,Holweck2012}) that there exist two inequivalent types of tripartite entanglement: the $GHZ$-type, which has a non-zero three-tangle and all two-tangles zero, and the $W$-type that has a vanishing three-tangle but the entanglement is balanced over all three parties. 
We find that our edges feature exclusively a $GHZ$-type tripartite entanglement; the other cases correspond either to completely separable states, or states showing two-partite entanglement not over all three pairs of parties. A detailed proof of this statement is given in the Appendix. It is of some interest to ask and have a look at how many `$GHZ$-entangled edges' are exhibited by different types of both kinds of magic configurations. The results of our computations are given in Table 4 (pentagrams) and Table 5 (WA-configurations).

\begin{table}[pth!]
\begin{center}
\footnotesize
\begin{tabular}{||l|c|r|r|r||}
\hline \hline
Hyperplane & Type & Pents &  String $[g_0,g_1,g_2,g_3,g_4,g_5]$ \\
\hline
\hline
trivial & $1$& $108$ & $[-,-,-,$54$,-,$54$]$      \\
 &$2$& $4104$ & $[-,$810$,$972$,$1836$,$324$,$162$]$      \\
 &$3$& $7884$ & $[${\it 648}$,$2862$,$2916$,$1134$,$324$,-]$      \\
\hline
$\mathcal{V}_{12}$ & $1$& $1$ & $[-,-,-,-,-,1]$      \\
 &$2$& $5$ & $[-,-,-,5,-,-]$      \\
 &$3$& $18$ & $[-,4,8,6,-,-]$      \\
\hline
$\mathcal{V}_{24}$ & $1$& $2$ & $[-,-,-,2,-,-]$      \\
 &$2$& $6$ & $[-,6,-,-,-,-]$      \\
 &$3$& $-$ &    $-$  \\
\hline
$\mathcal{V}_{20}$ & $1$& $-$ & $-$      \\
 &$2$& $-$ &    $-$  \\
 &$3$& $20$ &   $[-,20,-,-,-,-]$   \\
\hline
$\mathcal{V}_{14}$& $1$& $-$ & $-$      \\
 &$2$& $2$ &   $[-,2,-,-,-,-]$   \\
 &$3$& $14$ &  $ [-,6,8,-,-,-] $  \\
\hline
$\mathcal{V}_{3}$ & $1$& $2$ &$[-,-,-,-,-,2]$      \\
 &$2$& $84$ &  $ [-,24,24,36,-,-] $  \\
 &$3$& $250$ &   $[{\it 24},106,96,24,-,-]$   \\
\hline
$\mathcal{V}_{16}$ & $1$& $-$ & $-$      \\
 &$2$& $6$ &  $ [-,4,2,-,-,-] $  \\
 &$3$& $26$ &   $[4,16,6,-,-,-]$   \\
\hline
$\mathcal{V}_{21}$ & $1$& $-$ & $-$      \\
 &$2$& $12$ &  $ [-,6,4,2,-,-] $  \\
 &$3$& $24$ &   $[-,22,-,2,-,-]$   \\ 
\hline
$\mathcal{V}_{15}$ & $1$& $2$ &$[-,-,-,-,-,2]$      \\
 &$2$& $22$ &  $ [-,2,6,14,-,-] $  \\
 &$3$& $72$ &   $[{\it 4},14,36,16,2,-]$   \\
\hline
$\mathcal{V}_{22}$ & $1$& $-$ & $-$      \\
 &$2$& $-$ &    $-$  \\
 &$3$& $48$ &   $[-,26,22,-,-,-]$   \\
\hline
$\mathcal{V}_{22}$ ($2$-nd copy)& $1$ & $-$ &    $-$  \\
&$2$& $26$ &  $[-,-,14,6,2,-]$   \\
 &$3$& $22$ &   $[-,4,4,16,2,-]$   \\
\hline
$\mathcal{V}_{10}$& $1$& $-$ & $-$      \\
 &$2$& $36$ &   $[-,8,18,8,2,-]$   \\
 &$3$& $124$ &  $ [{\it 28},48,46,-,2,-] $  \\
\hline
$\mathcal{V}_{9}$ & $1$& $8$ &$[-,-,-,4,-,4]$      \\
 &$2$& $152$ &  $ [-,28,30,74,10,10] $  \\
 &$3$& $176$ &   $[{\it 4},34,90,38,10,-]$   \\
\hline
$\mathcal{V}_{5}$ & $1$& $16$ &$[-,-,-,8,-,8]$      \\
 &$2$& $136$ &  $ [-,16,16,76,16,12] $  \\
 &$3$& $280$ &   $[{\it 8},36,124,92,20,-]$   \\
\hline
$\mathcal{V}_{4}$ & $1$& $32$ &$[-,-,-,12,-,20]$      \\
 &$2$& $546$ &  $ [-,36,96,324,60,30] $  \\
 &$3$& ~~$878$ &   ~~$[-,146,372,264,96,-]$   \\
\hline \hline
\end{tabular}
\label{table4}
\caption{A distribution of pentagrams (within  the type of geometric hyperplane and across their particular type) in dependence on the number of `entangled' edges.  The numbers in italics correspond to pentagrams (always of type 3) devoid of entangled edges. 
}
\end{center}
\end{table}

Let us first discuss pentagrams. The entry $g_i$ ($i = 0, 1, 2, \ldots, 5$) in the string  $[g_0,g_1,g_2,g_3,g_4,g_5]$ stands for the number of pentagrams each of which possesses $i$ entangled edges. Thus, for example, a copy of $\mathcal{V}_{12}$ has one pentagram of type 1 whose all five edges are entangled, 5 pentagrams of type 2  where each has three edges showing entanglement and 18 pentagrams of type 3 of which four feature a single entangled edge, eight a couple of entangled edges and, finally,  six are endowed with three entangled edges each. From Table 4 it readily follows that pentagrams of type 1 feature entanglement on only three or five edges. 
Pentagrams of type 2 are more diversified, having always at least one of their edges entangled and offering some instances where all the five edges are entangled ($\mathcal{V}_9$, $\mathcal{V}_5$ and $\mathcal{V}_4$). Pentagrams of type 3 are even more variegated, with the possibility that there is no entangled edge (the case of hyperplanes $\mathcal{V}_3$, $\mathcal{V}_{15}$, $\mathcal{V}_{10}$, $\mathcal{V}_9$ and $\mathcal{V}_5$); there are altogether 648 distinct `unentangled' pentagrams. Interestingly, there is no string with all entries being non-zero, but several with just one entry non-zero. An example of $\mathcal{V}_{22}$ illustrates that strings are subject to change as we pass to a different copy of the hyperplane. Finally, we give an explicit factorization of the set of pentagrams in terms of the number of entangled edges,
\begin{equation}
12096=216(3+17+18+14+3+1),
\end{equation}
which is to be compared with eq.\,(1).

As per WA-configurations, the main result is that the maximum number of entangled edges is eight (out of 12 possible). One further sees
that out of four different types of WA's, only the last two feature no entangled edges (these being found in $\mathcal{V}_{22}$ and $\mathcal{V}_{10}$), and only types 2 and 3 contain configurations with the saturated upper bound ($\mathcal{V}_{5}$). Note that the last two types also share the $g_7 = 0$ property. Similarly to the previous case, there is no string with all entries being non-zero, but several ones where just one entry differs from zero. For the reader's convenience, we explicitly display the unique WA-configuration of type 3 that belongs to the selected copy of $\mathcal{V}_5$ and that has eight entangled edges (underlined) 
\begin{eqnarray*}
&[\underline{(IXY,XXI,XIY)},(XIY,YZX,ZZZ), (ZZZ,ZYX,IXY)],\nonumber \\
&[(ZIZ,ZZI,IZZ), \underline{(IZZ,XYX,XXY)}, \underline{(XXY,YXX,ZIZ)}],\nonumber \\
&[\underline{(ZXX,YYI,XZX)}, \underline{(XZX,ZXI,YYX)}, \underline{(YYX,XZI,ZXX)}],\nonumber \\
&[(YYI,ZZI,XXI), \underline{(ZXI,XYX,YZX)}, \underline{(XZI,YXX,ZYX)}],\nonumber 
\end{eqnarray*}
where the last three triples have index $-I$; this configuration is also remarkable by being uniquely extendible into this particular copy of  $\mathcal{V}_5$ (see \cite{Sanigaetal2012} for more details on this topic).

\begin{table}[t]
\begin{center}
\small
\begin{tabular}{||l|c|r|r|r||}
\hline \hline
Hyperplane & Type & WAs &  $[g_0,g_1,g_2,g_3,g_4,g_5,g_6,g_7;g_8]$  \\
\hline
\hline
$\mathcal{V}_{22}$ & $1$& $-$ & $-$      \\
 &$2$& $-$ &    $-$  \\
 &$3$& $1$ &   $[1,-,-,-,-,-,-,-;-]$   \\
 &$4$& $6$ &   $[3,-,-,3,-,-,-,-;-]$   \\
\hline
 $\mathcal{V}_{22}$ ($2$-nd copy)& $1$& $-$ & $-$      \\
 &$2$& $5$ &    $[-,-,1,-,1,2,1,-;-]$  \\
 &$3$& $1$ &   $[-,-,-,1,-,-,-,-;-]$   \\
 &$4$& $1$ &   $[-,1,-,-,-,-,-,-;-]$   \\
\hline
$\mathcal{V}_{10}$& $1$& $-$ & $-$      \\
 &$2$& $-$ &   $-$  \\
 &$3$& $4$ &   $[-,2,1,1,-,-,-,-;-]$   \\
 &$4$& $12$ &   $[1,5,3,2,1,-,-,-;-]$   \\
\hline
$\mathcal{V}_{9}$& $1$& $-$ & $-$      \\
 &$2$& $22$ &   $[-,-,4,6,2,4,5,1;-]$  \\
 &$3$& $16$ &   $[-,2,3,7,4,-,-,-;-]$   \\
 &$4$& $2$ &   $[-,-,-,2,-,-,-,-;-]$   \\
\hline
$\mathcal{V}_{5}$& $1$& $14$ &$[-,2,2,4,2,4,-,-;-]$       \\
 &$2$& $64$ &   $[-,-,6,13,13,6,19,4;{\it 3}]$  \\
 &$3$& $42$ &   $[-,2,10,7,18,2,2,-;{\it 1}]$   \\
 &$4$& $6$ &   $[-,3,-,-,1,-,2,-;-]$   \\
\hline
$\mathcal{V}_{4}$& $1$&$60$ &$[-,-,-,-,18,18,24;-]$       \\
 &$2$& $204$ &   $[-,-,12,12,48,72,30,30;-]$  \\
 &$3$& $60$ &   $[-,-,12,18,18,12,-,-;-]$   \\
 &$4$& $12$ &   $[-,-,-,-,6,6,-,-;-]$   \\
\hline \hline
\end{tabular}
\label{table5}
\caption{The same as in Table 4 for WA-configurations. However, unlike the previous case, the italicized numbers now denote configurations featuring the maximum number of entangled edges.  
}
\end{center}
\end{table}

\section{A few notable configurations of pentagrams}

The $108$ pentagrams of type 1 (see eq.\,(1)) comprise $36$ points and $81$ lines of $W(5,2)$, which can be viewed as a configuration of type $(36_{\{3,11\}},81_4)$, i.\,e. with $4$ points on a line and either $3$ or $11$ lines through a point.
Let us create a graph whose points are these 108 pentagrams and whose edges are pairs of mutually disjoint pentagrams. Its automorphism group is found to be isomorphic to $\mathbb{Z}_3^3 \rtimes D_6$, where $D_6$ is the $12$-element dihedral group.

Another noteworthy configuration stems from the $216$ `maximally entangled' pentagrams (see eq.\,(2), last term). The totality of their points and lines form a $(54_{\{24,32\}},378_4)$-configuration, whose automorphism graph is isomorphic to $\mathbb{Z}_3^3 \rtimes G_{48}$, where $G_{48}=\mathbb{Z}_2 \times S_4$.

Finally, the set of $27$ three-qubit operators that do not contain the identity matrix gives birth to a  $(27_8,54_4)$-configuration. This configuration underlies all $54$ `maximally-entangled' pentagrams of type 1 and its symmetry is that of the previous case.

\section{Conclusion}
We have carried out a detailed computer-based, finite-geometric analysis of two distinguished sets of three-qubit observables that serve as archetypal operator proofs 
of the Kochen-Specker theorem; a magic pentagram {\it \`a la} Mermin and a magic $(18_{2}, 12_{3})$ Waegell-Aravind configuration. It was found that
the symplectic polar space $W(5,2)$ contains altogether 12096 pentagrams. These fall into three different types according as the the number of edges whose observables multiply to $-I$ is five, three or one. Since 12096 is also the order of the automorphism group of the smallest split Cayley hexagon, we employed the latter geometry --- when classically embedded into $W(5,2)$ --- and looked at a distribution of pentagrams and their types within various types of hexagon's geometric hyperplanes. The results obtained seem to indicate that there is something deeper behind the above-mentioned numerical coincidence. The same procedure was then applied to the WA-configurations, of which there are four types according as the number of their edges indexed by $-I$ is seven, five, three or one. Here, again, the geometry of the hexagon sheds some intriguing light on the nature of these configurations. The main numerical findings are collected in two tables (Table 2 and 3). 

As each edge, of both a pentagram and a WA-configuration, features a set of  mutually commuting operators, it can also be associated with a set of eigenvectors common to the operators in question. Hence, a natural question arose: What kind of entanglement are such vectors endowed  with?
We found out that our edges feature exclusively a $GHZ$-type of entanglement; the other cases correspond either to completely separable states, or states showing two-partite entanglement not over all three pairs of parties.  It was, therefore, of considerable interest to ask about and have a computer look at how many `$GHZ$-entangled edges' are exhibited by different types of both kinds of magic configurations, either of their own or when situated in a randomly-chosen copy of a geometric hyperplane of the hexagon. A wealth of numerical results in this respect are again presented in tabular forms (Tables 4 and 5). 

The found rich finite-geometrical structure underlying both kinds of `magic' configurations may also be used as a novel material to feed unceasing discussions about the nature of non-locality and contextuality in quantum mechanics, and/or for designs  of experiments. Our results, when combined with those of \cite{Sanigaetal2012} and, to a lesser extent, also \cite{Levay2008}, establish a sort of relationship between Einstein's `elements of physical reality' and a certain `exceptional geometry' of compatible measurements, which is the smallest split Cayley hexagon. 

\section*{Acknowledgements}
This work was partially supported by the VEGA grant agency project 2/0098/10. We thank Dr. P. Vrana for enlightening correspondence concerning a skew symplectic embedding of the split Cayley hexagon and Dr. P. Pracna for providing us with the electronic versions of the figures.

\newpage
\section*{Appendix}
We shall show here why no edge of either a pentagram or a WA-configuration can feature a $W$-type of entanglement.
In fact, we have the more general following statement:\\

 {\em Assume that $u_1$ and $u_2$ are two commuting operators of the generalized three-qubit
 Pauli group and let $v$ be a common eigenvector, 
then $v$ is either non-entangled, partially entangled or exhibits entanglement of a GHZ-type}. \\
The proof can be carried out  in two steps:
\subsection*{Operators involving the identity at least once}
Let $u=U_1U_2U_3$ be an element of the generalized Pauli group where $U_i\in\{I, X, Y,Z\}$ and not all the three entries are simultaneously equal to the identity.
There are $36$ elements with at least one of the $U_i$ being equal to $I$. Assume $U_1=I$  and denote
 by $E_u ^\lambda$ the eigenspace of the operator $u$ for the eigenvalue $\lambda$. Then  
the  space of three qubits decomposes as $E_u^1\oplus E_u^{-1}$ and bases of the eigenspaces  are given by :

\begin{center}
\begin{tabular}{c|c}
$E_u^1$& $E_u^{-1}$\\
\hline
 $|000\rangle+|0U_2(|0\rangle)U_3(|0\rangle)\rangle$ &  $|000\rangle-|0U_2(|0\rangle)U_3(|0\rangle)\rangle$\\
 $|100\rangle+|1U_2(|0\rangle)U_3(|0\rangle)\rangle$ & $|100\rangle-|1U_2(0\rangle)U_3(|0\rangle)\rangle$\\
 $|01U_3(|1\rangle)\rangle+|0U_2(|1\rangle)1\rangle$&  $|01U_3(|1\rangle)\rangle-|0U_2(|1\rangle)1\rangle$\\
 $|11U_3(|1\rangle)\rangle+|1U_2(|1\rangle)1\rangle$ & $|11U_3(|1\rangle)\rangle-|1U_2(|1\rangle)1\rangle$\\
\end{tabular}
\end{center}

An eigenvector $v$ can therefore be written as 
$$v=(\alpha|0\rangle+\beta |1\rangle)\otimes (|00\rangle\pm |U_2(|0\rangle)U_3(|0\rangle)\rangle)+(\gamma|0\rangle+\delta |1\rangle)\otimes (|1U_3(|1\rangle)\rangle\pm |U_2(|1\rangle)1\rangle)$$
with $\alpha,\beta, \gamma, \delta\in \mathbb{C}$.
Such a vector is either non entangled, partially entangled or of GHZ-type. 
The same reasoning works if we consider $u=U_1IU_3$ or $u=U_1U_2I$.


\subsection*{Operators which do not involve the identity}

There are $27$ operators $u_i=U^i _1U^i _2U^i _3$ such that $U^i_j\in \{X,Y,Z\}$. 
Let $u_1=U_1 ^1 U_2^1U_3^1$ and $u_2=U_1^2U_2^2 U_3^2$ two such operators which commute. Because of the commuting assumption we necessarly have a $j\in \{1,2,3\}$ 
such that $U_j^1=U_j^2$. Without loss of generality let us assume that $j=1$ and let $v$ be a common eigenvector of $u_1$ and $u_2$.
The vector $v$ is also an eigenvector for $u=u_1u_2$. But by composition 
$u=u_1u_2=(U_1^1\times U_1^2)\otimes (U_2^1\times U_2^2)\otimes (U_3^1\times U_3^2)$, and  $U_1^1\times U_1 ^2=I$ by 
hypothesis (the notation $\times$ stands for the usual product of matrices).
Therefore $u=u_1u_2=U_2^3U_3^3$ where $U_2^3= U_2^1\times U_2^2$ and $U_3^3=U_3^1\times U_3^2$. 
In other words $u=u_1u_2$ is a Pauli operator of the three qubits system involving once the identity. 
By the previous step of our argument 
its eigenvector $v$ is either non-entangled, partially entangled or of GHZ-type.
�



\end{document}